\documentstyle[12pt]{article}
\begin{document}

\newcommand\beq{\begin{equation}}
\newcommand\eeq{\end{equation}}
\newcommand\bea{\begin{eqnarray}}
\newcommand\eea{\end{eqnarray}}

\centerline{\bf Quantization of the Derivative Nonlinear Schr\"{o}dinger
Equation}
\vskip 1 true cm

\centerline{Diptiman Sen \footnote{Electronic address: 
diptiman@physun.physics.mcmaster.ca}}
\small
\centerline{ \it Department of Physics and Astronomy, McMaster University,}
\centerline{ \it Hamilton, Ontario L8S 4M1, Canada}
\centerline{\it and Centre for Theoretical Studies, Indian Institute of 
Science,}
\centerline{\it Bangalore 560012, India}
\normalsize
\vskip 2 true cm

\noindent 
{\bf Abstract}
\vskip .5 true cm

We study the quantum mechanics of the derivative nonlinear Schr\"{o}dinger
equation which has appeared in many areas of physics and is known to be 
classically integrable. We find that the $N$-body quantum problem is exactly 
solvable with both bound states (with an upper bound on the particle number) 
and scattering states. Quantization provides an alternative way to understand 
various features of the classical model, such as chiral solitons and 
two-soliton scattering.
\vskip 1 true cm

 
\newpage

The derivative nonlinear Schr\"{o}dinger equation (DNLS) in one dimension
has historically found applications in many areas of physics, one example 
being circularly polarized nonlinear Alfv\'{e}n waves in a plasma \cite{WAD}. 
Related models have recently received fresh attention in the context of chiral 
Luttinger liquids; some of these models can be obtained by a 
dimensional reduction of a Chern-Simons model defined in two dimensions
[2-5]. The DNLS has some peculiarities, for instance, 
it is not Galilean invariant, and it has classical solitons which have an 
upper bound on the particle number and are chiral (with a particular sign 
of the momentum). In these respects, the DNLS differs from the usual nonlinear 
Schr\"{o}dinger equation (NLS), although both of them are classically 
integrable (see Refs. \cite{KAU,CLA} and references therein). Unlike the DNLS, 
the usual NLS has been studied semiclassically and 
quantum mechanically in great detail \cite{NOH,KUL}; being equivalent to a 
Bose gas with a two-body attractive $\delta$-function interaction 
\cite{LIE,MCG}. These studies have led to an alternative understanding 
of various features of the classical NLS, such as the solitons and their 
scattering. 

It therefore seems interesting to similarly analyze a quantum 
mechanical version of the DNLS. We do so in this Letter generalizing
the analysis of the two-body problem given in Refs. [2-5]. 
Besides providing us with a new exactly solvable $N$-body quantum mechanical 
problem, our study leads to a different and perhaps simpler way of deriving 
various properties of the DNLS, such as the peculiar features of the solitons
and the time delay in the scattering of two solitons.

We begin by considering a general Lagrangian density of the form
\beq
{\cal L} = i \hbar \psi^\star \partial_t \psi ~-~ \frac{\hbar^2}{2m} ~[~
\partial_x \psi^\star \partial_x \psi + i \lambda \rho ( \psi^\star \partial_x 
\psi - \partial_x \psi^\star \psi ) + 2 \mu \rho^2 + \nu \rho^3 ~]~ ,
\label{lag}
\eeq
where $\rho = \psi^\star \psi$ is the density with dimensions of inverse
length. We have introduced Planck's constant $\hbar$ 
in Eq. (\ref{lag}) for later convenience, although we will first discuss 
classical mechanics. The usual NLS is obtained by putting $\lambda = \nu =0$
in (\ref{lag}), while $\mu = \nu =0$ produces the DNLS. Note that the 
$\lambda$ term is not invariant under Galilean transformations 
\cite{AGL,JAC1}, and it flips sign under parity ($x \rightarrow - x$). It is 
therefore sufficient to consider the case $\lambda \ge 0$. The parameters 
$\lambda$ and $\nu$ are dimensionless while $\mu$ has the dimensions of 
inverse length; the system is scale invariant if we set $\mu =0$.

We will only examine the case $\nu = 0$ for the following reasons.
Firstly, it is known that the Euler-Lagrange equations 
of motion which follow from Eq. (\ref{lag}) are classically integrable
if $\nu = 0$, regardless of the values of $\lambda$ and $\mu$ \cite{KAU,CLA}.
Secondly, it is not clear to us how to handle the $\nu \rho^3$ term in 
(\ref{lag}) quantum mechanically. This piece would lead to a term like 
$\rho^2 \Psi$ 
in the Schr\"{o}dinger equation, where $\Psi$ is the wave function. If the
positions of the particles are denoted by $x_1, x_2, \ldots , x_N$, the density
operator is given by $\rho (x_i) = \sum_j \delta (x_i - x_j)$. Hence $\rho^2$
contains highly singular terms like $\delta^2 (x_i -x_j)$ as well as 
three-body terms like $\delta (x_i - x_j) \delta (x_i - x_k)$. (A possible
way of interpreting $\delta^2 (x_i - x_j)$  is as the limit of a two-body 
$\delta$-function interaction with infinite strength; this forces the wave 
function to vanish whenever the coordinates of two particles coincide 
\cite{AGL}. However the problem then becomes that of hard core bosons or free
fermions in one dimension which is easily solvable). We can avoid such 
singular interactions by setting $\nu =0$. This gives us the DNLS with a 
slight generalization if $\mu$ is nonzero.

For simplicity, we will set $\mu =0$ in Eq. (\ref{lag}) to start with. Later 
we will reintroduce $\mu$ and will comment on an interesting phenomenon which 
occurs in that case. We now briefly summarize the classical one-soliton 
solution of (\ref{lag}). The equation of motion is
\beq
i \hbar ~\partial_t \psi ~=~ - \frac{\hbar^2}{2m} ~\partial_x^2 \psi ~+~ 
\frac{i 2 \hbar^2 \lambda}{m} ~\rho ~\partial_x \psi ~.
\eeq
The density satisfies the continuity equation $\partial_t \rho + \partial_x j 
= 0$, where
\beq
j ~=~ -~ \frac{i \hbar}{2m} ~(\psi^\star \partial_x \psi - \partial_x 
\psi^\star \psi ) - \frac{\hbar \lambda}{m} \rho^2 ~.
\eeq
Since the system is integrable, there are an infinite number of conserved 
quantities \cite{KAU}. Three of these are the particle number, momentum
and energy,
\bea
N ~&=&~ \int ~dx ~\rho ~, \nonumber \\
P ~&=&~ - ~\frac{i \hbar}{2} ~\int ~dx ~ (\psi^\star \partial_x \psi - 
\partial_x \psi^\star \psi ) ~, \nonumber \\
E ~&=&~ \frac{\hbar^2}{2m} ~\int ~dx ~[~ \partial_x \psi^\star \partial_x \psi 
+ i \lambda \rho (\psi^\star \partial_x \psi - \partial_x \psi^\star \psi ) ~
]~ . 
\eea
With $\lambda > 0$, the one-soliton solution takes the form \cite{MIN,KAU}
\bea
\psi (x,t) ~&=&~ {\sqrt {\rho (x,t)}} ~\exp ~[ ~i~ (kx - \omega t + \lambda 
\int_{-\infty}^x dy ~\rho (y,t) ~) ~]~, \nonumber \\
\rho (x,t) ~&=&~ \frac{\alpha^2}{2 \lambda (~{\sqrt{\alpha^2 + 4k^2}} ~\cosh 
\alpha (x-vt) ~+~ 2k)} ~, \nonumber \\
k ~&=&~ \frac{mv}{\hbar} \quad {\rm and} \quad  \omega ~=~ \frac{mv^2}{2 
\hbar} - \frac{\hbar \alpha^2}{8m} ~.
\label{sol}
\eea
These expressions contain two independent parameters, the velocity $v$ and the 
inverse width of the soliton $\alpha$. On computing the three conserved 
quantities, we find that $0 < N \lambda < \pi$, and
\bea
\cos N \lambda ~&=&~ \frac{2k}{{\sqrt {\alpha^2 + 4k^2}}} ~, 
\nonumber \\
M ~&=&~ \frac{m}{\lambda} ~\tan N \lambda ~,
\label{mcl}
\eea
with $P =  Mv$ and $E = P^2 /2M$.
We observe that $P$ is always positive, although $M$, $v$ and $E$ are positive
if $0< N \lambda < \pi /2$ and negative if $\pi /2 < \cos N \lambda < 
\pi$. ($P$ would have been negative if we had chosen $\lambda < 0$).

We now study this system quantum mechanically, and show that the classical 
picture is recovered in the limit $N \rightarrow \infty$. The Schr\"{o}dinger 
equation describing $N$ identical bosons is easily derived \cite{AGL,MIN} 
after defining the wave function
\beq
\Psi (x_1, x_2, \ldots , x_N, t) ~=~ \langle 0 \vert {\hat \Psi} (x_1, t) 
{\hat \Psi} (x_2, t) \ldots {\hat \Psi} (x_N, t) \vert N \rangle ~,
\eeq
where $\hat \Psi (x,t)$ denotes the second-quantized bosonic field operator. 
We then obtain the equation
\beq
i \hbar ~\partial_t \Psi ~=~ - \frac{\hbar^2}{2m} ~\sum_i ~\partial_{x_i}^2 
\Psi ~+~ \frac{i \hbar^2 \lambda}{m} ~\sum_{i<j} ~\delta (x_i - x_j ) 
(\partial_{x_i} + \partial_{x_j} ) \Psi ~,
\label{sch}
\eeq
where we have normal ordered the Hamiltonian to eliminate divergent 
self-interaction terms like $\delta (x_i - x_i) = \delta (0)$. We thus find a
two-body $\delta$-function interaction whose strength depends on the total 
momentum of the two particles. Note that we do not need to worry about the 
relative ordering of $\partial_{x_i} + \partial_{x_j}$ and $\delta (x_i - 
x_j )$ since they commute; thus the $\delta$-function interaction does not 
affect the total momentum of the two particles.

Before proceeding further, it is instructive to examine the two-particle 
problem. For the configuration $x_1 < x_2$, let us assume a stationary solution
of the form
\beq
\Psi ~=~  \exp ~[~ \frac{i}{\hbar} (c_1 x_1 + c_2 x_2)~] ~+~ a ~\exp ~[~ 
\frac{i}{\hbar} (c_2 x_1 + c_1 x_2)~] ~,
\label{psi1}
\eeq
where $c_1$ and $c_2$ may be complex, and $a$ will be called the reflection
amplitude. The wave function for $x_2 < x_1$ is then obtained by Bose symmetry.
For convenience, we will refer to the exponentials in Eq. (\ref{psi1}) as waves
and the $c_n$'s as the particle momenta even if the $c_n$'s are not real.
The total momentum of (\ref{psi1}) is $P = c_1 + c_2$. We can now go to the 
center of mass and relative coordinates $X = (x_1 + x_2)/2$ and $x = x_1 - 
x_2$, and factor out the center of mass wave function. The relative motion is 
described by the equation
\beq
i \hbar \partial_t \Psi ~=~ [~ - ~\frac{\hbar^2}{m} \partial_x^2 ~-~ 
\frac{\hbar^2 \lambda P}{m} \delta (x) ~+~ \frac{P^2}{4m} ~] ~\Psi ~.
\eeq
We then find that the amplitude $a$ in Eq. (\ref{psi1}) is given by 
\beq
a ~=~ \frac{c_2 - c_1 - i \lambda ( c_2 + c_1)}{c_2 - c_1 + i \lambda ( c_2 + 
c_1)} ~.
\label{ref}
\eeq
Clearly, there is a bound state if $a=0$ and $P > 0$.

We now seek a bound state solution of the $N$-particle Schr\"{o}dinger
equation (\ref{sch}),
\beq
\Psi ~=~ \exp ~[~ \frac{i}{\hbar} ~\sum_n ~c_n x_n ~] 
\label{bou}
\eeq
for the configuration $x_1 < x_2 < \ldots < x_N$; the wave function for all 
other configurations can be deduced by Bose symmetry. We find that (\ref{bou}) 
satisfies the boundary conditions of the $\delta$-function interactions if the 
reflection amplitude given in Eq. (\ref{ref}) is zero for each neighboring 
pair of particles, i.e., 
\beq
c_{n+1} - c_n ~=~ i \lambda ~ (c_{n+1} + c_n ) 
\label{cn1}
\eeq
for $n=1,2, \ldots , N-1$. (It is because of the absence of the reflected 
waves that our wave function (\ref{bou}) has only one wave, instead of $N!$ as 
in the general Bethe ansatz). We now demand that the total momentum should 
be equal to $P = \sum_n c_n$, and discover that the $c_n$'s are given by
\beq
c_n ~=~ P ~\frac{\sin \theta}{\sin N \theta} ~\exp ~[~ i (2n-N-1) \theta ~]~,
\label{cn2}
\eeq
where
\beq
\exp ~(i 2 \theta ) ~=~ \frac{1 + i \lambda}{1 - i \lambda} ~.
\eeq
The parameter $\theta$ lies in the range $[0, \pi /2]$ for $\lambda 
> 0$. The energy is given by $E = \sum_n c_n^2 /2m = P^2 /2M$, where
\beq
M ~=~ m ~\frac{\tan N \theta}{\tan \theta} ~.
\label{mqu}
\eeq

We now ask, when do Eqs. (\ref{bou}) and (\ref{cn2}) describe a bound state, 
i.e., a state which is normalizable if we use translation invariance to fix 
the centre of mass at some particular place? It is easy to show that the 
necessary and sufficient condition for this is that 
\beq
P ~\frac{\sin \theta}{\sin N \theta} ~\sum_{n=1}^{l} ~\sin (N+1-2n) \theta ~
=~ P ~\frac{\sin (N-l) \theta ~ \sin l \theta}{\sin N \theta} ~>~ 0 ~,
\label{cond}
\eeq
for all values of $l$ from $1$ to $N/2$ if $N$ is even and to $(N-1)/2$ if $N$
is odd. The conditions in (\ref{cond}) arise on demanding that the probability
$\Psi^\star \Psi$ should go to zero if we take the $l$ particles on the right 
($x_{N+1-l}, x_{N+2-l}, \ldots , x_N$) to $\infty$ and the $l$ particles on the
left ($x_1, x_2, \ldots , x_l$) to $- \infty$ (thereby keeping the centre of 
mass
fixed). For small values of $N$, we can check whether or not Eqs. (\ref{cond}) 
are satisfied; we find an intricate pattern of allowed values of $\theta$ and 
$P$ (which sometimes has to be negative) for which a bound state exists. The 
situation simplifies in the limit $N \rightarrow \infty$ and 
$\theta \rightarrow 0$ keeping $N \theta$ fixed. We then find a bound state
if and only if $N \theta < \pi$ and $P > 0$. Further, in this limit, 
$\theta = \lambda$ so that the classical and quantum formulae (\ref{mcl}) 
and (\ref{mqu}) for the masses agree; we therefore identify the quantum
bound state with the classical soliton. (It would be interesting to derive
the soliton profile in Eq. (\ref{sol}) from the wave function in (\ref{bou})
using a technique given in Ref. \cite{NOH}). We thus see that only a finite 
number of particles can be bound for a given value of $\theta$. It is quite 
remarkable that if $\pi / 2 < N \theta < \pi$, the energy of the bound state 
can be lowered arbitrarily by giving it sufficiently large momentum.

Next we study the scattering of bound states. As a warmup exercise, 
consider the scattering of a bound object of $N-1$ particles with momentum
$P-p$ with one particle of momentum $p$, where $p$ is real. For $x_1 < \ldots <
x_N$, we find that the wave function can be written as a superposition of $N$
waves,
\bea
\Psi = && \exp ~[ \frac{i}{\hbar} (c_1 x_1 + \ldots + c_{N-1} x_{N-1} + p 
x_N )] 
\nonumber \\
&& + A \exp ~[ \frac{i}{\hbar} (p x_1 + c_1 x_2 + \ldots + c_{N-1} x_N )] 
\nonumber \\
&& + \sum_{n=1}^{N-2} ~a_n \exp ~[ \frac{i}{\hbar} (c_1 x_1 + \ldots + c_n 
x_n + p x_{n+1} + \ldots + c_{N-1} x_N ) ],
\label{psi2}
\eea
where the $c_n$'s are given by Eq. (\ref{cn2}) with $N$ and $P$ replaced by 
$N-1$ and $P-p$ respectively in that equation. The first two waves in 
(\ref{psi2}) correspond
respectively to configurations in which the the bound state is entirely to 
the left and entirely to the right of the free particle; $A$ is therefore the 
transmission amplitude for the particle to go through the bound object. We 
now use Eq. (\ref{ref}) repeatedly to relate all the $a_n$'s and $A$ to each 
other. $A$ is found to be a pure phase of the form $A = C/C^\star$, where
\beq
C ~=~ \prod_{n=1}^{N-1} ~\Bigl( ~\frac{p}{\sin \theta} ~e^{-i \theta} ~-~ 
\frac{P-p}{\sin (N-1) \theta} ~e^{i (2n-N+1) \theta } ~\Bigr) ~.
\eeq

We now examine the general scattering of two bound objects (solitons), and
again discover that there is only transmission and no reflection. We 
consider one object with $N_1$ particles and momentum $P_1$ and another 
object with $N_2$ particles and momentum $P_2$; let $N_1+N_2 =N$. 
We introduce $N_1$ numbers $c_n^{(1)}$ satisfying Eq. (\ref{cn2}) 
(with $N$, $P$ replaced by $N_1$, $P_1$), and $N_2$ numbers $c_n^{(2)}$ 
satisfying (\ref{cn2}) with $N_2$, $P_2$. For a given configuration $x_1 < 
x_2 < \ldots < x_N$, we then find that the wave 
function is a superposition of several waves, each wave having the particle 
momenta as some permutation of the numbers $c_n^{(1)}$ and $c_n^{(2)}$. We do
not get all the $N!$ possible permutations due to the absence of reflection 
within the momenta $c_n^{(1)}$'s, or within the $c_n^{(2)}$'s, as
discussed around Eq. (\ref{cn1}). The permutations allowed are those in which
particles $i$ and $j$ can have the momenta $c_n^{(1)}$ and $c_{n+1}^{(1)}$ (or
momenta $c_n^{(2)}$ and $c_{n+1}^{(2)}$) only if $i < j$. The number of
such permutations is $N!/N_1 ! N_2 !$. Two of these describe configurations in
which the object with $N_1$ particles is entirely to the left (or to the 
right) of the object with $N_2$ particles; all the other permutations 
describe configurations in which the $N_1$ particles are interspersed 
amongst the $N_2$ particles. Thus the wave function has the form 
\bea
\Psi ~=& &~ \exp ~[~ \frac{i}{\hbar} (\sum_{n=1}^{N_1} ~c_n^{(1)} x_n + 
\sum_{n=1}^{N_2} c_n^{(2)} x_{N_1 +n}) ~] \nonumber \\
& &+~ A~ ~\exp ~[~ \frac{i}{\hbar} (\sum_{n=1}^{N_2} ~c_n^{(2)} x_n + 
\sum_{n=1}^{N_1} c_n^{(1)} x_{N_2 +n}) ~] \nonumber \\
& &+~ \quad {\rm all ~the ~other ~waves} ~,
\eea
where $A$ denotes the transmission amplitude for one bound state to pass 
through the other. We compute $A$ by successively passing each of the 
$N_2$ particles on the right through each of the $N_1$ particles on the left,
and using the expressions in Eq. (\ref{ref}) at each such crossing. We finally 
discover that $A = \exp (i2 \delta ) =C/ C^\star$, where
\bea
C ~=~ & & \prod_{n=2}^{N_2} ~\Bigl( ~\frac{P_2}{\sin N_2 \theta} ~e^{i(2n-2-
N_2) \theta} ~ -~ \frac{P_1}{\sin N_1 \theta} ~e^{i N_1 \theta} ~\Bigr) 
\nonumber \\
& \cdot & \prod_{n=1}^{N_1} ~\Bigl( ~\frac{P_2}{\sin N_2 \theta} ~e^{-iN_2 
\theta} ~-~ \frac{P_1}{\sin N_1 \theta} ~e^{i(2n-N_1) \theta} ~\Bigr) ~.
\eea
Let us consider the case in which the velocities $v_i = (2E_i/M_i)^{1/2}$ 
satisfy $v_1 > v_2 > 0$. For weak coupling ($\theta \rightarrow 0$ with $N_1$, 
$N_2$ held fixed, so that $M_i = m N_i$), we find the phase shift
\beq
2 \delta ~=~ 2 \theta N_1 N_2 ~\frac{{\sqrt{E_1/M_1}} + {\sqrt{E_2/M_2}}}{
{\sqrt{E_1/M_1}} - {\sqrt{E_2/M_2}}} ~+~ O(\theta^2) ~.
\label{pha}
\eeq
From this, we can compute the time delay suffered by the bound state with 
energy $E_1$; thus $\Delta t_1 = 2 \hbar \partial \delta /\partial E_1$ 
\cite{JAC2}. The expression in (\ref{pha}) and the resultant time delay 
$\Delta t_1$ agree with the results in Ref. \cite{MIN} if we set $N_1 = N_2$.

Let us now include the term proportional to $\mu$ in Eq. (\ref{lag}). This 
adds $(2 \hbar^2 \mu /m) \sum_{i<j} \delta (x_i - x_j )$ to the right hand 
side of the Schr\"{o}dinger equation (\ref{sch}). For a bound state of
$N$ particles with momentum $P$, we find that the numbers $c_n$ in 
Eq. (\ref{bou}) are given by
\beq
c_n ~=~ \Bigl(P- \frac{N \hbar \mu}{\tan \theta} \Bigr)~ \frac{\sin 
\theta}{\sin N \theta} ~\exp ~[~i (2n-N-1) \theta ~] ~+~ \frac{\hbar \mu}{\tan 
\theta} ~.
\label{cn3}
\eeq
As before, we find that in the limit $\theta \rightarrow 0$ with $N \theta$ 
held fixed, the bound state exists only if $N \theta < \pi$; in addition, we
need $P > N \hbar \mu / \tan \theta$. More interestingly, we observe that the
energy $E = \sum_n c_n^2 /2m$ is minimum at a {\it nonzero} value of the 
momentum given by
\beq
P_0 ~=~ \frac{N \hbar \mu}{\tan \theta} ~\Bigl( ~1 ~-~ \frac{\tan N \theta}{N 
\tan \theta} ~\Bigr) 
\eeq
if $\mu \tan N \theta < 0$. For weak coupling ($N \theta < \pi /2$), we can 
understand this result as follows. We know that there is a zero momentum 
soliton if $\mu < 0$ (attractive interaction) even if $\theta =0$ 
\cite{NOH,MCG}; the energy of this is given by 
\beq
E ~=~ - ~\frac{\hbar^2 \mu^2}{6m} ~(N^3 - N) ~,
\label{ene}
\eeq
as we can see by setting $P=0$ and taking the limit $\theta \rightarrow 0$
in Eq. (\ref{cn3}). Now if $\theta$ is small and positive, the strength of the 
attractive $\delta$-function interaction is increased if all the particles 
move with positive momentum; this lowers the energy by effectively increasing 
the value of $\mu^2$ in (\ref{ene}). Thus zero momentum is not the state of 
lowest energy if $\theta$ is nonzero and $\mu < 0$.

For completeness, we would like to mention the states in which all the 
particles have real momenta $p_n$. (However these purely scattering states 
are not the lowest energy states of our system). It is convenient to put the 
system on a circle with circumference $L$, and consider a particular ordering 
of the positions $0 < x_1 < \ldots < x_N < L$. The wave function is then 
given by the general Bethe ansatz with a superposition of $N!$ waves. 
Following Ref. \cite{LIE}, we impose periodic boundary conditions
\beq
\Psi (0,x_2, \ldots ,x_N) ~=~ \Psi (x_2, \ldots ,x_N,L) ~.
\eeq
We then find that the $p_n$'s are related to each other by the $N$ equations
\beq
\frac{p_n L}{\hbar} ~=~ 2\pi I_n ~+~ \pi (N-1) ~+~ 2~\sum_{l=1}^N ~ 
\tan^{-1} ~ \Bigl( ~\frac{p_n -p_l}{\lambda (p_n +p_l) -2 \hbar \mu} ~\Bigr) ~,
\eeq
where the $I_n$'s are integers. These equations may be solved numerically. 

Finally, we ask whether the system has a well-defined ground state in the 
thermodynamic limit $N, L 
\rightarrow \infty$ keeping $N/L$ fixed. Our earlier analysis indicates that 
the answer is no, even if $\mu$ is zero or even positive (repulsive). We 
have seen that a bound state with $N$ 
particles can have an arbitrarily low energy if its momentum is large and 
$\pi / 2 < N \theta < \pi$. If $N$ is very large, the system can lower 
its energy arbitrarily by forming a number of large momentum bound objects 
with particle numbers $N_1, N_2, \ldots , N_k$ (adding upto $N$), such that 
$\pi /2 < N_i \theta < \pi$ for $i=1,2, \ldots ,k$.

To conclude, the quantization of the DNLS has produced a rich structure. It
would be interesting to consider other classically integrable systems and see
if they can be quantized in order to shed new light on them.

\vskip .5 true cm
I thank Rajat Bhaduri for discussions and the Department of Physics and 
Astronomy, McMaster University for its hospitality during the course of 
this work. This research was supported by the Natural Sciences and Engineering
Research Council of Canada.

\vskip .5 true cm
\noindent {\it Note Added:}

After writing this paper, I learnt that similar work has been published 
earlier \cite{SHN,LAI}; I thank A. G. Shnirman for pointing this out. 

\vskip 1 true cm

\end{document}